\documentclass[10pt,twocolumn]{article}

\usepackage[utf8]{inputenc}
\usepackage[T1]{fontenc}

\usepackage{graphicx}

\usepackage{lipsum}
\usepackage{xcolor}

\usepackage{authblk}
\usepackage[margin=2cm]{geometry}
\usepackage{mathtools, cuted}

\usepackage{bm}
\usepackage{amsfonts}
\usepackage{amsmath}
\usepackage{amssymb}
\usepackage{mathrsfs} 
\usepackage{siunitx}
\usepackage{pdfpages}

\title{Evaporation-induced temperature gradient in a foam column}
\author[1]{François Boulogne}
\author[1]{Emmanuelle Rio}
\author[1]{Frédéric Restagno}
\affil[1]{Université Paris-Saclay, CNRS, Laboratoire de Physique des Solides, 91405, Orsay, France.}

\date{\today}
\begin{document}

\twocolumn[
    \begin{@twocolumnfalse}
        \maketitle
        \begin{abstract}
            Various parameters affect the foam stability: surface and bulk rheology of the solution, gravitational drainage, mechanical vibrations, bubble gas composition, and also evaporation.
            Evaporation is often considered through the prism of a liquid loss, but it also induces a cooling effect due to the enthalpy of vaporization.
            In this study, we combine a theoretical and experimental approach to explore the temperature field in a foam column evaporating from the top.
            We show that a measurable temperature profile exists in this geometry with temperatures at the interface lower than the environmental temperature by few degrees.
            We demonstrate that the temperature profile is the result of a balance between the enthalpy of vaporization and heat fluxes originating from the thermal conduction of foam and air and the thermal radiation.
            For small foam thicknesses compared to the radius, we found that the temperature gradient is established over the foam thickness while for large aspect ratios, the gradient is spanning over a lengthscale comparable to the tube radius.
        \end{abstract}
        \vspace{2em}
    \end{@twocolumnfalse}
]

\section{Introduction}

Foam stability \cite{Langevin2018} is one of the crucial parameter for successful applications \cite{Hill2017}.
In such an assembly of bubbly, the liquid tend to flow downward due to gravity (drainage) \cite{Cohen2013}, the small bubbles tend to empty in the bigger ones (coarsening) \cite{Saint2006,Stevenson2010}, neighboring bubbles tend to merge (coalescence) \cite{Rio2014}, and liquid evaporation affects the foam wetness.
These different mechanisms at the origin of foam aging are either entangled, coarsening and drainage for example, or poorly understood like coalescence or evaporation, which has been mainly ignored in the literature. 
To build an empirical knowledge, systematic experimental tests have been developed to compare chemical formulations in various physical and chemical conditions (pH, temperature, etc.).
A common approach is to use foam columns and measure the temporal evolution of the foam height \cite{Pugh2016,Wang2016}.
For instance, some commercial instruments rely on the Bikerman test measuring the equilibrium height in presence of a given bubbling velocity \cite{Bikerman1938} or on the Ross-Milles test based on the time evolution of the height of a foam generated beforehand \cite{Ross1941}.

Most of the time, these tests are performed in closed atmosphere.
However, in ambient conditions, the liquid evaporation plays also a role on the stability of soap films.
This aspect has recently received particular attention on soap films, based on the consideration that evaporation induces thinning of films \cite{Champougny2018,Poulain2018,Pasquet2022}.
On foams, the effect of evaporation has also been noticed by several authors \cite{Pertsov1980,ekserova_foam_1998,Stevenson2006,Tobin2011}.
Li \textit{et al.}  suggested that the film thinning comes with an increase of the surfactant concentration where evaporation takes place \cite{Li2012a}.
Such concentration gradient leads to a Marangoni flow increasing further the film thinning rate, and thus promoting the film bursting. 
The opposite effect, \textit{i.e.} a stability enhanced by evaporation due to Marangoni effects has been obtained by Chandran Suja \textit{et al.} on oil-based systems \cite{ChandranSuja2018}.
As a result, laboratory tests performed with sealed containers may not reflect the same behavior as in the practical conditions of usage.
Therefore, the effect of evaporation on foams must be carefully considered.

More recently, we have evidenced that the temperature of a soap film is not always equal to the ambient temperature due to evaporation \cite{Boulogne2022}.
The cooling effect appears to be significant, up to $8$~$^\circ$C, for a film of 12~mm of diameter and a relative humidity of about 20~\%.
To the best of our knowledge, the cooling effect induced by evaporation is overlooked in foams, which may have important consequences to better understand and predict their  stability.

In this paper, we propose to investigate experimentally and theoretically the cooling induced by the evaporation of a foam column.
We measure the temperature profile in foams of different aspect ratios.
Then, we rationalize these measurements by predicting the steady-state temperature of the foam-air interface and the resulting temperature profile in the foam column.
Finally, we discuss the predictions offered by the model and we compare the predicted temperature profiles to our measurements.

\section{Experimental procedure and observations}

\begin{figure}
    \centering
    \includegraphics[width=.6\linewidth]{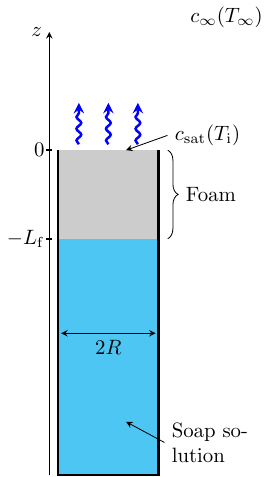}
    \caption{Foam column of height $L_{\rm f}$ and radius $R$ evaporating at the rim, $z=0$, in the atmosphere characterized by the vapor concentration $c_\infty$ at the temperature $T_\infty$.}
    \label{fig:setup}
\end{figure}

Figure \ref{fig:setup} is a schematic of the experiment.
The column of foam is made in a cylindrical tube of radius $R=17$~mm and of 20~cm total height.
Tubes are in PMMA (Abaqueplast) with a wall thickness of $3$~mm.
A soap solution prepared by mixing a commercial dish washing soap (Fairy) at 10 \%wt with pure water (resistivity = 18.2 M$\Omega \cdot$cm), is poured in the tube such that the liquid interface reaches a distance $L_{\rm f}>0$ to the rim.
The surface tension of the liquid is $\gamma = 25.4 \pm 0.1$~mN/m.
The foam is produced by bubbling air with a pressure generator (OF1, Elveflow, France) through a 32G blunt needle in the soap solution \cite{Drenckhan2015}.
This method produces a monodisperse foam with a bubble diameter of $2b=2$~mm; the bubble size is determined by image analysis of the rising bubbles \cite{Marchand2020}.
The choice of a monodisperse foam is motivated by the limited coarsening process \cite{Drenckhan2010} and to the seek of a homogeneous, easy to reproduce material.
As illustrated in figure~\ref{fig:setup}, the foam is generated up to the rim and is placed in an environment controlled in relative humidity at ${\cal R}_{\rm H} = 50$~\% \cite{Boulogne2019}.
The environmental temperature  is measured for each experiment and is $T_\infty = 21\pm1$~$^\circ$C.

The temperature is measured by a thermocouple probe (type K, NiAl-NiCr, diameter 0.2~mm, RS PRO) connected to a digital thermometer (RS PRO 1314).
The probe is directly plunged in the foam.
The position of the probe is controlled by a motorized translation stage (LTS150, Thorlabs) and the temperature is recorded once stabilized at each position.
Temperatures are measured with a typical uncertainty of $\pm 0.1$~$^\circ$C and the positions with a typical uncertainty of 1~mm.
The relative humidity is measured with a HIH-4021-003 sensor from Honeywell, USA.

\begin{figure}
    \centering
    \includegraphics[width=.99\linewidth]{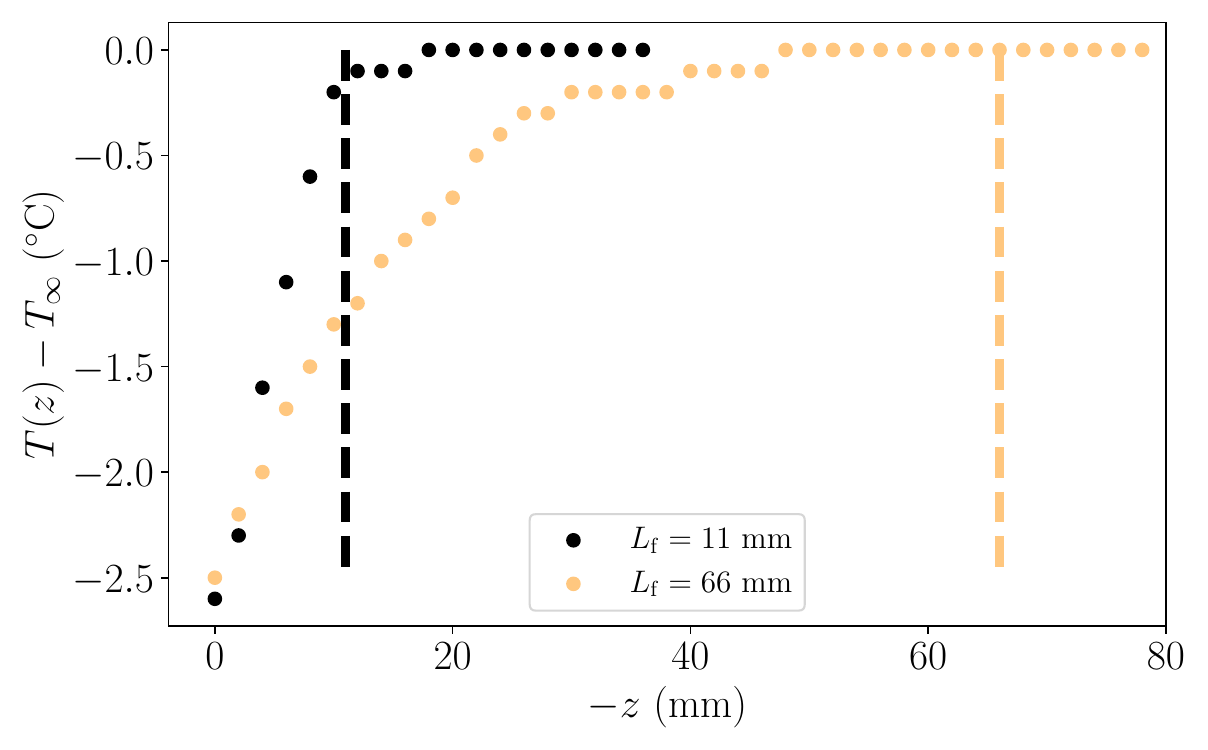}
    \caption{Measurement of the temperature profile for two foam thicknesses at a relative humidity ${\cal R}_{\rm H}=48$~\%.
    These thicknesses are materialized by vertical dashed lines at $-z=L_{\rm f} =11$ and 66~mm.
    The column radius is $17$~mm, such that the ratios $L_{\rm f}/R$ are 0.6 and 3.9, respectively.}
    \label{fig:results}
\end{figure}

Typical measurements are presented in figure~\ref{fig:results} where we calculate  the difference of temperature between the thermocouple at a given position $T(z)$ and the ambient temperature $T_\infty$ measured with the same instrument.
We show two temperature profiles for foam thicknesses smaller and larger than the tube radius, \textit{i.e.} $L_{\rm f}/R = 0.6$ and $L_{\rm f}/R = 3.9$ respectively.
Within the duration of the experiments (typically 10 minutes), we have not noticed a significant foam aging or bubble bursting that would have changed the typical thickness of the foam layer or the bubble size.

The temperatures of the foam/atmosphere interface are nearly equal in both situations, to the resolution of the temperature measurements and the determination of the position of the interface ($\pm$ 1 mm).
This surface temperature is $2.5$~$^\circ$C lower than the environmental temperature for both foam heights.
However, the characteristic lengthscale of the temperature variation differs with the foam height.
For an aspect ratio $L_{\rm f}/R\simeq 0.6$, the temperature nearly reaches the ambient temperature at the foam-liquid interface, which results in a sharp temperature increases with the penetration in the foam.
In contrast, for the thicker foam with an aspect ratio $L_{\rm f}/R\simeq 3.9$, the temperature variation is smoother and the ambient temperature is reached ahead of the liquid-foam interface.

Additionally, we checked by weighing samples in identical conditions that the soap solution evaporates at the same rate as pure water, indicating that the solutes do not alter significantly the chemical potential of the solution \cite{Boulogne2022}.

In the next section, we aim at predicting the temperature profile in the foam, which includes the determination of the temperature of the interface due to evaporation.

\section{Model}

\subsection{Evaporative flux}

The atmosphere is characterized by its temperature $T_\infty$ and it relative humidity $\mathcal{R}_{\rm H}$, which is defined as $p_\infty / p_{\rm sat}(T_\infty) $ where $p_\infty$ is the partial vapor pressure at $T_\infty$ and $p_{\rm sat}(T_\infty)$, the saturated vapor pressure of water.
The temperature variation of the saturated vapor pressure is well described by the Antoine equation $p_{\rm sat}(T) = p^\circ\, 10^{A - B/(C + T)}$ where $p^\circ = 10^5$~Pa and the coefficients where $A = 5.341 \pm 0.003$~K, $B = 1807.5 \pm 1.6$~K, and $C = -33.9 \pm 0.1$~K are fitted with data from~\cite{Rankin2009}.

In a steady state regime, assuming diffusive transfers between the foam and the atmosphere, the vapor concentration field is the solution of the Laplace equation.
The total evaporative flux $Q_{\rm ev}$ of a circular disk of radius $R$ is proportional to the difference of vapor concentrations $\Delta c^\star$ between the environment at a temperature $T_\infty$ and above the interface at a temperature $T_{\rm i}$.
This evaporative flux writes
\begin{equation}\label{eq:Q_ev}
    Q_{\rm ev} = 4{\cal D} R \Delta c^\star,
\end{equation}
where ${\cal D}$ is the diffusion coefficient of vapor in air \cite{Cooke1967,Lebedev1965}.

The difference of vapor concentrations $\Delta c^\star = c_\infty - c_{\rm sat} (T_{\rm i})$ can be related to the difference of vapor pressures.
Denoting $P$ the atmospheric pressure, $\rho_{\rm air}$ the air density, $M_{\ell}$ and $M_{\rm air}$ the molar weight of the liquid and air respectively, we have $\Delta c^\star \simeq \frac{\rho_{\rm air} M_{\ell}}{M_{\rm air}} \frac{\Delta p^\star}{P}$ with $\Delta p^\star =  p_\infty - p_{\rm sat} (T_{\rm i})$.

Due to the enthalpy of vaporization, the temperature of the interface $T_{\rm i}$ is lower than the ambient temperature.
Thus, the interface receives a heat flux from the environment.
Three contributions can be identified: by conduction from the surrounding atmosphere, by conduction from the foam, and also by radiation.
In the following paragraphs, we model these heat transfers.

\subsection{Heat fluxes}

\paragraph{Heat flux from the atmosphere}
Similarly to the vapor concentration field, the temperature field is the solution of a Laplace equation with $T_{\rm i}$ at the interface and $T_\infty$ far from it.
Thus, the heat flux from the atmosphere is
\begin{equation}
    Q_{\rm h1} = 4 \lambda_{\rm air} R \Delta T^\star,
\end{equation}
where   $\lambda_{\rm air} = 0.025$ W$\cdot$K$^{-1}\cdot$m$^{-1}$ is the thermal conductivity of air and $\Delta T^\star = T_\infty - T_{\rm i}$ is the temperature difference.
Next, to determine the heat flux from the foam, we must comment its thermal conductivity.

\paragraph{Thermal conductivity of foams}

The foam is supposed to be a continuous medium, which is valid if the lengthscale of the heterogeneities is much smaller than the size of the material, \textit{i.e.} for $b \ll \{R, L_{\rm f}\}$.
The foam and the soap solution are characterized by their thermal conductivity, $\lambda_{\rm f}$ and $\lambda_{\ell}$, respectively.
Leach  proposed a model combining series and parallel contributions to the thermal conductivity \cite{Leach1993}.
At the first leading order in $\varphi$, the liquid fraction, the thermal conductivity of foam can be estimated as

\begin{equation}\label{eq:thermal_conductivity_foam}
    \lambda_{\rm f} = \lambda_{\rm air} + \frac{2}{3}\lambda_{\ell}  \varphi,
\end{equation}
with $\lambda_{\ell} = 0.61$ W$\cdot$K$^{-1}\cdot$m$^{-1}$.

The liquid fraction of the foam results from the balance between gravity and capillary suction, which implies that the liquid fraction decreases with the altitude.
The liquid fraction profile $\varphi(z)$ is obtained by solving the drainage equation in a stationary regime (See Eq.~(3.89) in \cite{Cantat2013}), which gives
\begin{equation}\label{eq:phi_ell}
    \varphi(z) = \hat\varphi \left( \frac{L_{\rm f} + z}{\ell_c} + \left(\frac{\varphi_c}{\hat\varphi}\right)^{-1/2} \right)^{-2},
\end{equation}
where $\varphi_c = 0.26$ is the fraction of gaps in a close-packing of hard spheres and $\hat\varphi = \ell_c^2 / b^2\delta^2$ with the capillary length $\ell_c = \sqrt{\gamma / \rho g}$ and a geometric constant $\delta = 1.73$.

\begin{figure}
    \centering
    \includegraphics[width=\linewidth]{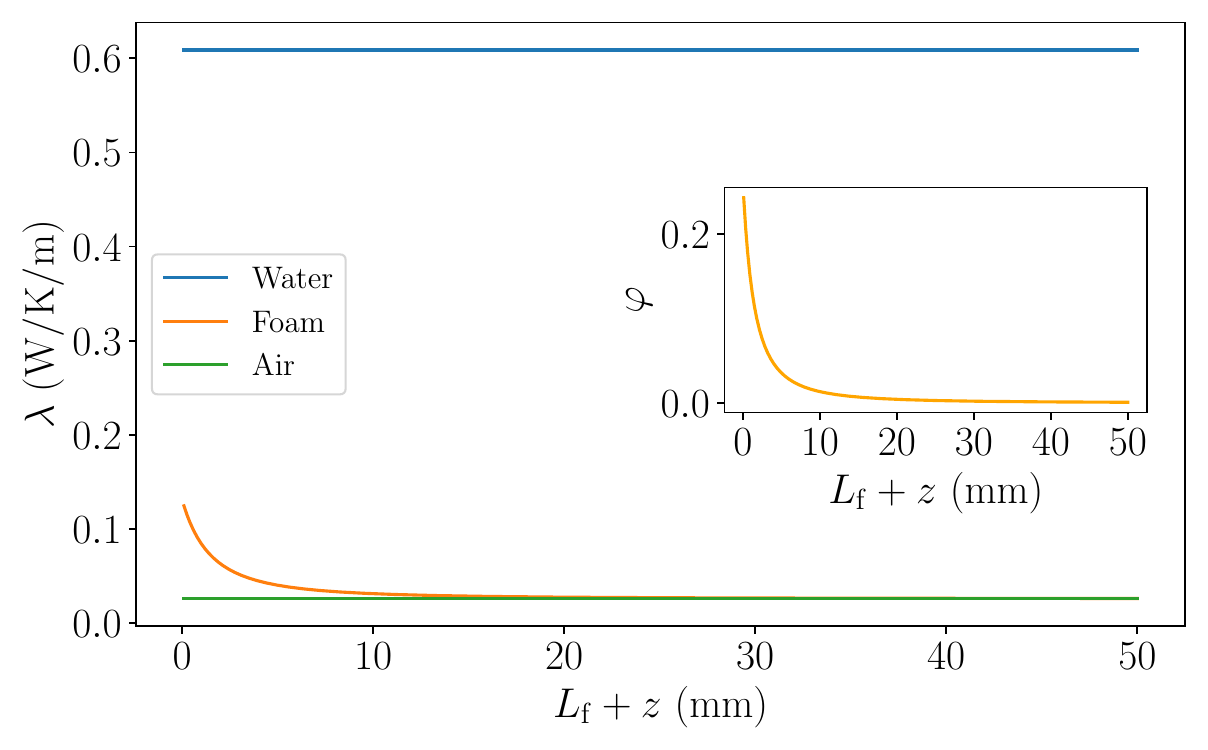}
    \caption{The main plot shows the thermal conductivity of the foam as the function of the altitude, computed from equation \ref{eq:thermal_conductivity_foam}.
    As an indication, the thermal conductivity values of water and air are also represented.
    In the inset, the liquid fraction is plotted as a function of the vertical position from equation \ref{eq:phi_ell}.
    Computations are performed for a bubble radius $b=1$~mm and  $\ell_c=1.6$~mm.
    As an indication, at $L_{\rm f} + z=25$~mm, $\lambda_{\rm f}/\lambda_{\rm air} = 1.04$.}
    \label{fig:thermal_conducivity}
\end{figure}

Equation \ref{eq:phi_ell} is plotted in the inset of figure~\ref{fig:thermal_conducivity} for a bubble radius $b=1$~mm.
A wet region located near the liquid-foam interface spans over a few bubble layers where a sharp decrease of the liquid fraction occurs, followed by a smoother decrease with the altitude \cite{Cantat2013,Maestro2013}.
From the liquid fraction profile, we plot in figure~\ref{fig:thermal_conducivity} the thermal conductivity as a function of the altitude $z$ with equation~\ref{eq:thermal_conductivity_foam}.
Again, we notice a sharp variation of $\lambda_{\rm f}$ over the first few layers of bubbles and the thermal conductivity converges quickly to the air conductivity.

In the following, we consider that the thermal conductivity of the foam is independent of the vertical position $z$ and we take $\lambda_{\rm f} \simeq \lambda_{\rm air}$.
This approximation allows to perform analytical calculations that will facilitate the interpretation of the predictions made by the model.
After a comparison of the predictions with experimental measurements, we will comment further this approximation in the discussion.

\paragraph{Heat flux from the foam column}

We propose to describe the temperature field in the center of the foam $T(z)$.
The environment is at a temperature $T_\infty$.
Due to the small thickness of the tube wall compared to the tube radius and the larger thermal conductivity of the wall compared to the foam, the temperature variation in the tube wall is neglected.
Thus, the temperature profile $T(z)$ at the center of the column results of the heat transfers from the column periphery, at $T_\infty$, the interface at $T_{\rm i}$, and the liquid, which is far from the liquid-foam interface at $T(z\rightarrow -\infty) = T_\infty$.

A heat flux balance on a slice between $z$ and $z + {\rm d}z$ yields the differential equation
\begin{equation} \label{eq:eq_diff_T_z}
 \frac{{\rm d}^2T}{{\rm d}z^2} + \frac{2}{R^2} \left(T_\infty - T(z)\right) = 0.
\end{equation}
This equation is valid both in the foam, which lies from $z=0$ to $z=-L_{\rm f}$, and in the liquid from $z=-L_{\rm f}$ to $z\rightarrow -\infty$.
The boundary conditions associated with the differential equation \ref{eq:eq_diff_T_z} are
\begin{subequations}\label{eq:BCs}
\begin{align}
    T(0) &= T_{\rm i},\label{eq:BC1}\\
    T(z\rightarrow -\infty) &= T_\infty,\label{eq:BC2}\\
    T\left(-L_{\rm f}^-\right) &= T\left(-L_{\rm f}^+\right),\label{eq:BC3}\\
    \lambda_{\rm f} \left.\frac{{\rm d} T}{{\rm d} z}\right|_{-L_{\rm f}^+} &= \lambda_{\rm \ell} \left.\frac{{\rm d} T}{{\rm d} z}\right|_{-L_{\rm f}^-}\label{eq:BC4},
\end{align}
\end{subequations}
where equations \ref{eq:BC1} and \ref{eq:BC2} are the temperature conditions at the interface and in the liquid far from the foam, respectively; equation  \ref{eq:BC3} corresponds to the temperature continuity at the liquid-foam interface, and \ref{eq:BC4} is the continuity of the thermal flux at the liquid-foam interface.

The solution of equation \ref{eq:eq_diff_T_z} is expressed in the foam and the liquid respectively, and reads

\begin{equation}
\label{eq:temp_field}
T_{\ell, \rm f}(z) =  T_\infty -\Delta T^\star \left[  \alpha_{\ell, \rm f} e^{-\sqrt{2} z/R} - \beta_{\ell, \rm f}  e^{\sqrt{2} z/R} \right],
\end{equation}
where the coefficients $\alpha_{\rm f}$, $\beta_{\rm f}$, $\alpha_{\ell}$, $\beta_{\ell}$ are determined with the boundary conditions \ref{eq:BCs}a-d.
Let us denote  $k^\pm = \exp(\pm \sqrt{2} L_{\rm f} / R)$ and $\Lambda = \lambda_{\rm f} / \lambda_{\ell}$.
Then, we have

\begin{subequations}\label{eq:temp_field_coeff}
\begin{align}
\alpha_{\ell} &= 0,\\
\beta_{\ell} &= \frac{k^+}{k^-} + \beta_{\rm f} \left(1 + \frac{k^+}{k^-} \right),\\
\alpha_{\rm f} &= 1 + \beta_{\rm f},\\
\beta_{\rm f} &= \frac{k^+ (1 + \Lambda)}{ k^-(1 - \Lambda) - k^+(1 + \Lambda)}.
\end{align}
\end{subequations}

\begin{figure}
    \centering
    \includegraphics[width=.99\linewidth]{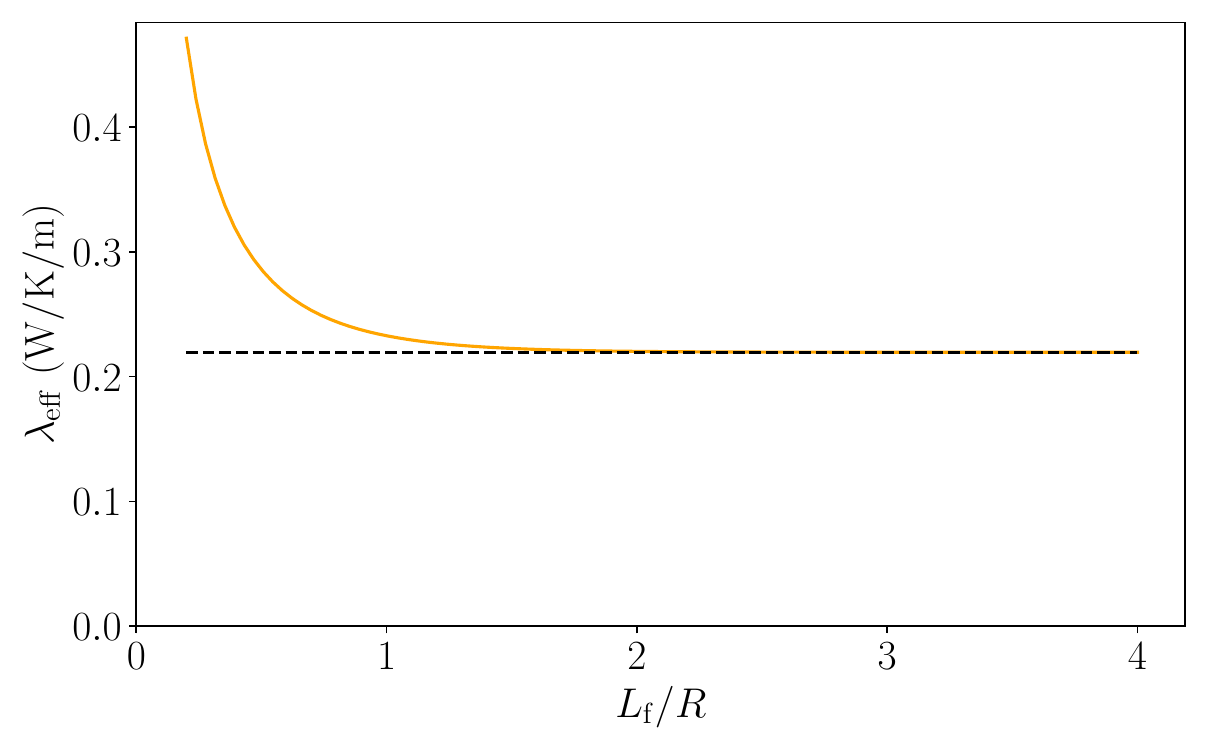}
    \caption{The effective thermal conductivity $\lambda_{\rm eff}$ as defined in equation \ref{eq:Q_h} is plotted as a function of the foam aspect ratio $L_{\rm f}/R$ in orange.
    The dashed black line is the limit for large aspect ratio
    $\lambda_{\rm eff}^{\rm lim} = 4 \lambda_{\rm air} + \pi \sqrt{2} \lambda_{\rm f}$.
    }
    \label{fig:lambda_eff}
\end{figure}

 The heat flux from the bulk of the foam to the interface is $Q_{\rm h2} = - \pi R^2 \lambda_{\rm f} \left. \overrightarrow{z} \cdot \overrightarrow{\nabla}T \right|_{z =0} $.
 Combined with equation \ref{eq:temp_field}, we obtain
\begin{equation}\label{eq:Q_h2}
     Q_{\rm h2} =  - \pi \sqrt{2} R \lambda_{\rm f} \left(1  + 2 \beta_{\rm f} \right) \Delta T^\star.
\end{equation}
It is convenient to define the total heat flux by conduction $Q_{\rm h} = Q_{\rm h1} + Q_{\rm h2}$, which can be written in the form
\begin{equation}\label{eq:Q_h}
    Q_{\rm h} = \lambda_{\rm eff} R \Delta T^\star,
\end{equation}
with an effective thermal conductivity $\lambda_{\rm eff} = 4 \lambda_{\rm air} - \pi \sqrt{2}  \left(1  + 2 \beta_{\rm f} \right) \lambda_{\rm f}$.
This effective thermal conductivity depends on foam aspect ratio $L_{\rm f}/R$ through the coefficient $\beta_{\rm f}$ as shown in figure~\ref{fig:lambda_eff}.
In the limit of large aspect ratios, we have $\beta_{\rm f}\rightarrow -1$.
Thus, the effective thermal conductivity becomes independent of the aspect ratio and the value is $\lambda_{\rm eff}^{\rm lim} = 4 \lambda_{\rm air} + \pi \sqrt{2} \lambda_{\rm f}$.
Both terms in the expression of $\lambda_{\rm eff}$ are of the same order of magnitude, which reflects the significance of both $Q_{\rm h1}$ and $Q_{\rm h2}$ in the heat exchange with the interface.

\paragraph{Radiative flux}
Radiation is known to be significant to estimate the cooling effect when the characteristic size of the evaporating surface is typically larger than several millimeters \cite{Hill1916,Boulogne2022}.
We describe the radiative flux bringing heat to the surface by the Stefan-Boltzmann equation
  \begin{equation}\label{eq:Q_rad}
    Q_{\rm rad} = \pi R^2 \epsilon \sigma (T_\infty^4 - T_{\rm i}^4),
\end{equation}
where $\epsilon$ is the emissivity and $\sigma$ the Stefan-Boltzmann constant.
Below, we will comment the significance of this radiative flux with respect to the conductive heat flux $Q_{\rm h}$.

Now that we determined the temperature profile in the column and the heat transfers, the temperature of the interface $T_{\rm i}$ remains to be calculated to close the problem.
In the next paragraph, we derive this temperature from the balance between the energy absorbed by the evaporation and the heat transfers.

\subsection{Energy balance}
In a steady state regime, the energy balance writes
\begin{equation}\label{eq:heat_balance}
    h_{\rm ev} Q_{\rm ev}  = - Q_h\left( 1 + \frac{  Q_{\rm rad}}{Q_h}\right),
\end{equation}
where $h_{\rm ev}$ is the enthalpy of vaporization.
For small temperature differences, \textit{i.e.} $|T_{\rm i} - T_\infty| / T_\infty\ll 1$, the ratio of radiative and conductive fluxes can be simplified as
\begin{equation}\label{eq:Q_rad_Q_h}
    \frac{Q_{\rm rad}}{Q_{\rm h}} = \frac{ \pi R \epsilon \sigma (T_\infty^4 - T_{\rm i}^4)}{\lambda_{\rm eff}  (T_\infty - T_{\rm i}) } \simeq \frac{4 \pi R \epsilon \sigma T_\infty^3}{\lambda_{\rm eff}}.
\end{equation}

Considering large foam aspect ratios, the radiative flux appears to be as significant as the conductive flux for a critical radius $R_{\rm c} = \lambda_{\rm eff}^{\rm lim} / (4 \pi \epsilon \sigma T_\infty^3)$.
At $21~^\text{o}$C, we find $R_{\rm c} \simeq 7$~mm, which indicates that $Q_{\rm rad}$ cannot be neglected in our experiments.

From equation \ref{eq:heat_balance}, with equations \ref{eq:Q_ev}, \ref{eq:Q_h}, and \ref{eq:Q_rad_Q_h}, we relate the difference of vapor pressures $\Delta p^\star$ to the temperature difference $\Delta T^\star$ that reads
\begin{equation}\label{eq:psychro}
     \Delta p^\star  = - P  {\cal A} \Delta T^\star,
\end{equation}
where ${\cal A}$ is the so-called psychrometer coefficient, which is for this system,
\begin{equation}\label{eq:psychro_coeff}
{\cal A} = \frac{M_{\rm air}   }{\rho_{\rm air} M_{\ell}  } \frac{\lambda_{\rm eff}}{4 h_{\rm ev} {\cal D}} \left( 1 + \frac{4 \pi R \epsilon \sigma T_\infty^3}{\lambda_{\rm eff}}\right).
\end{equation}

We can remark that $\Delta p^\star$ depends on the temperature of the interface $T_{\rm i}$, which is necessary to take into account the variation of the evaporative flux with the temperature of the interface.
The temperature dependence of $\Delta p^\star$ being non-linear due to Antoine equation, the interfacial temperature $T_{\rm i}$ cannot be determined analytically from equation \ref{eq:psychro}.

As a consequence, for a given column radius $R$, ambient temperature $T_\infty$, and relative humidity ${\cal R}_{\rm H} = ( p_{\rm sat}(T_{\rm i}) - \Delta p^\star) / p_{\rm sat}(T_\infty) $,  we seek numerically for the temperature $T_{\rm i}$ from equation \ref{eq:psychro} with the Antoine equation by using a Newton procedure \cite{Virtanen2020}.
Next, the temperature profile is computed with  equations \ref{eq:temp_field} and \ref{eq:temp_field_coeff}.

From equation \ref{eq:psychro_coeff}, it is worth noting that the psychrometer coefficient depends on lengthscales through (a) $\lambda_{\rm eff}$, which depends itself of $L_{\rm f}/R$ for small aspect ratios, and (b) directly of the tube radius $R$  due to the ratio  $Q_{\rm rad} / Q_{\rm h}$ (Eq.~\ref{eq:Q_rad_Q_h}).
Thus, the psychrometer coefficient is a function of $(R, L_{\rm f}/R)$.
Therefore, the temperature of the interface cannot be rescaled neither by the radius $R$, nor by the aspect ratio $L_{\rm f}/R$, and so the temperature profiles.

In the next section, we discuss the results of this model, and we compare the predictions to experiments.

\section{Discussion}\label{sec:discussion}

\begin{figure}
    \centering
    \includegraphics[width=.99\linewidth]{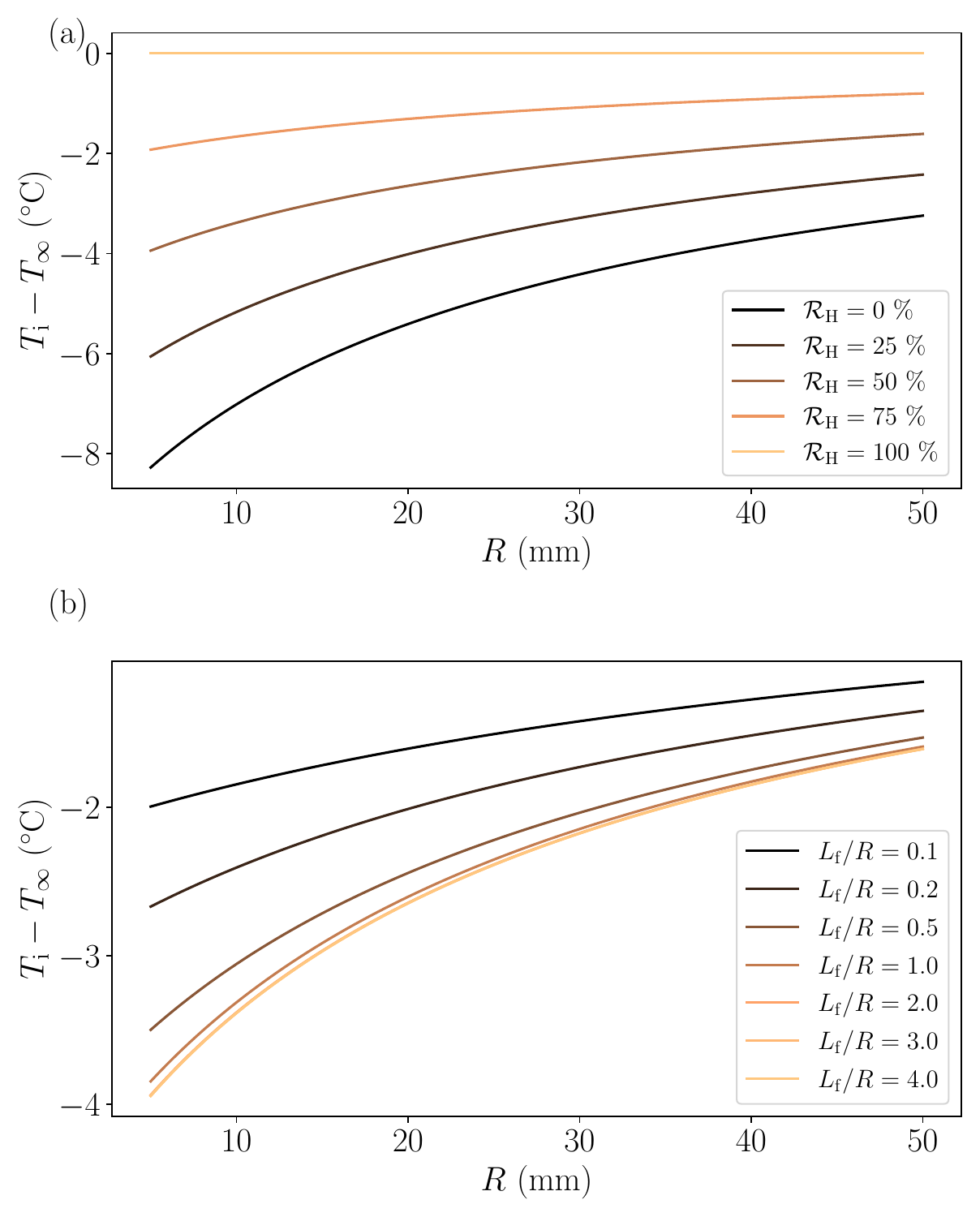}
    \caption{Theoretical prediction of the cooling effect on the interface as a function of the foam column radius.
    (a) Effect of the relative humidity for $L_{\rm f}/R = 3$.
    (b) Influence of the foam height with respect to the foam radius.
    The ambient temperature is set at $T_\infty=20$ $^\circ$C and ${\cal R}_{\rm H} = 50$~\%.
    }
    \label{fig:temperature_interface}
\end{figure}

Before comparing the predictions to the experiments, we start by analyzing the influence of the model parameters on the temperature of the interface.
In figure \ref{fig:temperature_interface}(a), we plot the cooling effect as a function of the foam column radius for different values of the relative humidity and a constant foam aspect ratio.

Evaporation and thus the cooling effect being driven by the relative humidity, a dry atmosphere leads to a lower temperature of the interface.
This statement is a direct consequence of the psychrometric equation (Eq.~\ref{eq:psychro}).
In addition, the increase of the foam column radius reduces the cooling effect.
Indeed, in figure \ref{fig:temperature_interface}(a), the ratio $Q_{\rm rad}/Q_{\rm h}$ increases linearly with the radius (Eq.~\ref{eq:Q_rad_Q_h}) such that wider columns receive a more important contribution of the radiative transfer, which hinders the cooling effect.

The foam aspect ratio also has an effect on the temperature of the interface.
In figure~\ref{fig:temperature_interface}(b), we plot the temperature difference as a function of the radius for aspect ratio smaller and larger than unity.
For small aspect ratios, typically $L_{\rm f}/R < 1$, the temperature of the interface decreases with the foam thickness at a given radius.
This observation is directly linked to the increased effective thermal conductivity $\lambda_{\rm eff}$ observed in figure~\ref{fig:lambda_eff}.
Schematically, a small foam thickness reduces the insulation of the foam-vapor interface with the underlying liquid, which has a significantly larger thermal conductivity than air.
Thus, increasing the aspect ratio beyond unity has a little effect on the interface temperature, as the temperature profile becomes dominated by the heat flux from the wall of the container.

\begin{figure}
    \centering
    \includegraphics[width=.99\linewidth]{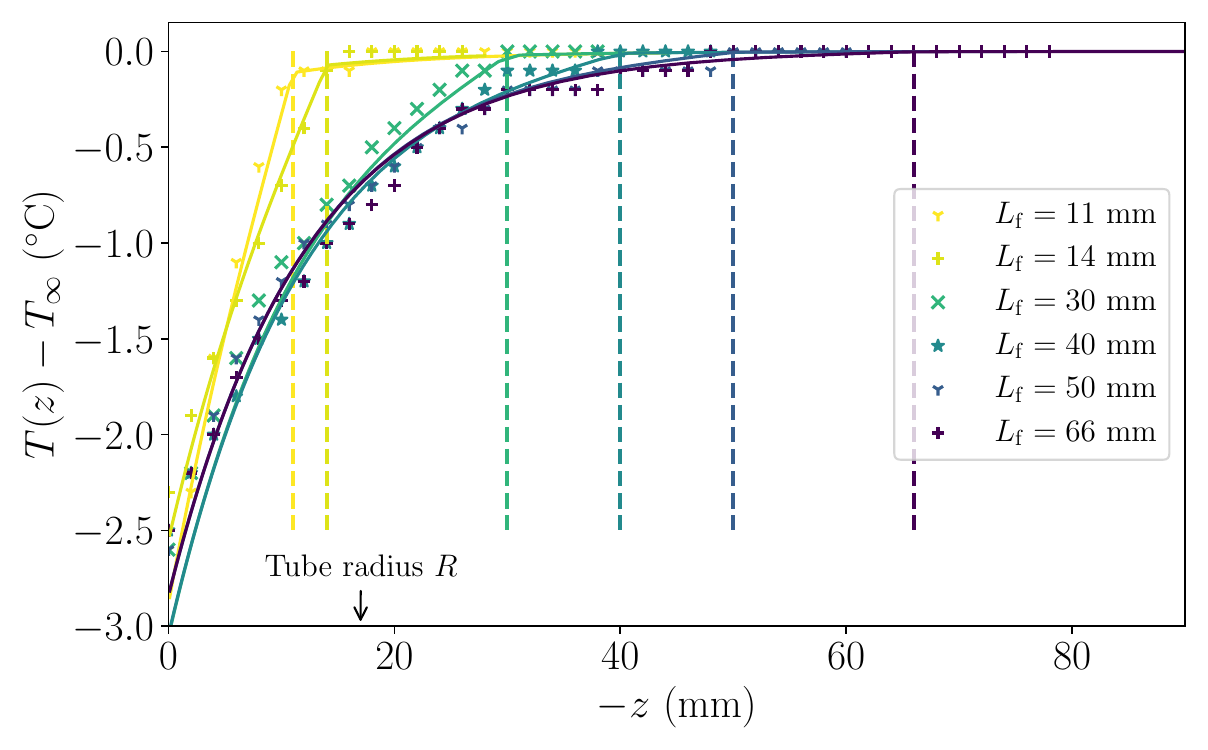}
    \caption{Comparison between the measurement of the temperature profile for various foam thicknesses with the prediction given by equation~\ref{eq:temp_field}.
    The experiments are performed in a column of radius  $R=17$~mm at a relative humidity ${\cal R}_{\rm H}=0.48$.
    The vertical dashed lines correspond to the positions of the liquid/foam interface $L_f$.
  }  \label{fig:results_interpreted}
\end{figure}

Now that the effects of the relative humidity and the size of the tube are clarified, we can focus on the temperature profile in the foam for different aspect ratios.
In figure~\ref{fig:results_interpreted}, we show the experimental measurements presented in figure~\ref{fig:results} with additional foam heights.
We also plot the temperature profiles given by equation~\ref{eq:temp_field} without any fitting parameter.
We observe that the model is in excellent agreement with the measurements, within the experimental uncertainties.

The temperature profiles presented in figure~\ref{fig:results_interpreted} clearly illustrate the insulating properties of foam, especially for small aspect ratios where a neat rupture in the profile is observed at the liquid-foam interface.
In this case, the lengthscale associated with the temperature gradient is the foam thickness.
Also, for larger aspect ratios, the profiles tend to collapse on a single curve.
Again, this collapse indicates that the temperature is mainly the results of the radial heat flux over the vertical flux.
Thus, for large foam aspect ratios, the characteristic lengthscale of the temperature gradient is the tube radius.

Finally, we come back to the approximation we made on the thermal conductivity of the liquid foam.
The assumption of a uniform thermal conductivity to describe the temperature profile in the foam columns is well validated by the experiments as shown in figure~\ref{fig:results_interpreted}.
Indeed, we expect that this approximation is less satisfactory when the temperature gradient, and thus the heat flux, at the liquid-foam interface becomes significant, which is the case for thin foam layers.
Temperatures of thinner foam layers must deviate from the present model, but this is difficult to render with our experimental protocol as the foam thickness will be composed of only a few bubble layers, conflicting with our continuum approach.

\section{Conclusions}

In this work, we have investigated experimentally and theoretically the temperature profile that is established in a foam column evaporating from the top.
Experimentally, we observed that a temperature profile is perfectly measurable with temperature differences between the interface and the ambient temperature of few degrees.
Considering the foam aspect ratio as the ratio between the foam thickness and the tube radius, we found for thin foams, the lengthscale of the temperature gradient is set by the foam thickness.
Conversely, for large aspect ratios, the lengthscale is the tube radius.

We successfully modeled the temperature profile in the column with an analytical expression by considering, in a 1D approximation, a heat flux balance between the radial and the axial directions, assuming a constant thermal conductivity in the foam layer.
The temperature of the interface results from an energy balance between the energy associated with the enthalpy of vaporization and the thermal flux from the environment.
The latter originates from three contributions: the heat conduction by the foam and by the surrounding air, as well as the radiative flux.
The good comparison of our theoretical predictions with our experiments validates the aforementioned approximations.

Further studies will be necessary to evaluate how the temperature gradient combined with the evaporation plays a role on the foam stability.
The foam drying due to evaporation promotes bubble bursting \cite{Carrier2003}, while the role of thermal gradients is more subtle.
For instance, evaporation can increase the concentration of non-volatile compounds, increasing also the bubble stability \cite{Roux2022}.
 a Marangoni flow in surface bubbles placed in a cool atmosphere has been found to enhance the stability by thickening the soap film \cite{Nath2022}, which remains an open question on foams.

\section*{Acknowledgments}
The authors thank A. Commereuc and M. Pasquet for stimulating discussions.
The authors acknowledge for funding support from the French Agence Nationale de la Recherche in the framework of project AsperFoam - 19-CE30-0002-01.

\bibliography{biblio}

\begin{thebibliography}{10}

\bibitem{Langevin2018}
D.~Langevin.
\newblock Foam stabilization mechanisms.
\newblock In {\em Foam Films and Foams}, pages 213--232. CRC Press, 2018.

\bibitem{Hill2017}
C.~Hill and J.~Eastoe.
\newblock Foams: From nature to industry.
\newblock {\em Adv. Colloid Interface Sci.}, 247:496--513, 2017.

\bibitem{Cohen2013}
S.~Cohen-Addad, R.~Höhler, and O.~Pitois.
\newblock Flow in foams and flowing foams.
\newblock {\em Annu. Rev. Fluid Mech.}, 45:241--267, 2013.

\bibitem{Saint2006}
A.~Saint-Jalmes.
\newblock Physical chemistry in foam drainage and coarsening.
\newblock {\em Soft Matter}, 2(10):836--849, 2006.

\bibitem{Stevenson2010}
P.~Stevenson.
\newblock Inter-bubble gas diffusion in liquid foam.
\newblock {\em Current Opinion in Colloid \& Interface Science},
  15(5):374--381, 2010.

\bibitem{Rio2014}
E.~Rio and A.-L. Biance.
\newblock Thermodynamic and mechanical timescales involved in foam film rupture
  and liquid foam coalescence.
\newblock {\em Chem. Phys. Chem.}, 15(17):3692--3707, 2014.

\bibitem{Pugh2016}
R.~J. Pugh.
\newblock {\em Bubble and foam chemistry}.
\newblock Cambridge University Press, 2016.

\bibitem{Wang2016}
J.~Wang, A.~V. Nguyen, and S.~Farrokhpay.
\newblock A critical review of the growth, drainage and collapse of foams.
\newblock {\em Adv. Colloid Interface Sci.}, 228:55--70, 2016.

\bibitem{Bikerman1938}
J.J. Bikerman.
\newblock The unit of foaminess.
\newblock {\em Trans. Faraday Soc.}, 34:634--638, 1938.

\bibitem{Ross1941}
J.~Ross and G.~D. Miles.
\newblock An apparatus for comparison of foaming properties of soaps and
  detergents.
\newblock {\em Oil \& Soap}, 18(5):99--102, 1941.

\bibitem{Champougny2018}
L.~Champougny, J.~Miguet, R.~Henaff, F.~Restagno, F.~Boulogne, and E.~Rio.
\newblock Influence of {Evaporation} on {Soap} {Film} {Rupture}.
\newblock {\em Langmuir}, 34(10):3221--3227, 2018.

\bibitem{Poulain2018}
S.~Poulain, E.~Villermaux, and L.~Bourouiba.
\newblock Ageing and burst of surface bubbles.
\newblock {\em J. Fluid Mech.}, 851:636–671, 2018.

\bibitem{Pasquet2022}
M.~Pasquet, F.~Boulogne, J.~Saint-Anna, F.~Restagno, and E.~Rio.
\newblock Impact of physical-chemistry on the film thinning in surface bubbles.
\newblock {\em Soft Matter}, 18:4536--4542, 2022.

\bibitem{Pertsov1980}
A.~V. Pertsov, V.~F. Borachuk, B'~E. Chlstyakov, and E.~D. Shchukin.
\newblock Evaporation of the dispersion medium and foam stability.
\newblock {\em Dokl. Akad. Nauk SSSR}, 1980.

\bibitem{ekserova_foam_1998}
D.~R. Ekserova and P.~M. Krugliakov.
\newblock {\em Foam and foam films: theory, experiment, application}.
\newblock Number vol. 5 in Studies in interface science. Elsevier, Amsterdam ;
  New York, 1998.

\bibitem{Stevenson2006}
P.~Stevenson.
\newblock The wetness of a rising foam.
\newblock {\em Ind. Eng. Chem. Res.}, 45(2):803--807, January 2006.

\bibitem{Tobin2011}
S.~T. Tobin, A.~J. Meagher, B.~Bulfin, M.~Möbius, and S.~Hutzler.
\newblock A public study of the lifetime distribution of soap films.
\newblock {\em Am. J. Phys.}, 79(8):819--824, August 2011.

\bibitem{Li2012a}
X.~Li, S.~I. Karakashev, G.~M. Evans, and P.~Stevenson.
\newblock Effect of environmental humidity on static foam stability.
\newblock {\em Langmuir}, 28(9):4060--4068, 2012.

\bibitem{ChandranSuja2018}
V.~Chandran~Suja, A.~Kar, W.~Cates, S.M. Remmert, P.D. Savage, and G.G. Fuller.
\newblock Evaporation-induced foam stabilization in lubricating oils.
\newblock {\em Proc. Natl. Acad. Sci.}, 115(31):7919--7924, 2018.

\bibitem{Boulogne2022}
F.~Boulogne, F.~Restagno, and E.~Rio.
\newblock Measurement of the temperature decrease in evaporating soap films.
\newblock {\em Phys. Rev. Lett.}, 129:268001, 2022.

\bibitem{Drenckhan2015}
W.~Drenckhan and A.~Saint-Jalmes.
\newblock The science of foaming.
\newblock {\em Adv. Colloid Interface Sci.}, 222:228 -- 259, 2015.

\bibitem{Marchand2020}
M.~Marchand, F.~Restagno, E.~Rio, and F.~Boulogne.
\newblock Roughness-induced friction in liquid foams.
\newblock {\em Phys. Rev. Lett.}, 124:118003, 2020.

\bibitem{Drenckhan2010}
W.~Drenckhan and D.~Langevin.
\newblock Monodisperse foams in one to three dimensions.
\newblock {\em Curr. Opin. Colloid Interface Sci.}, 15(5):341--358, 2010.

\bibitem{Boulogne2019}
F.~Boulogne.
\newblock Cheap and versatile humidity regulator for environmentally controlled
  experiments.
\newblock {\em Eur. Phys. J. E}, 42(4):51, 2019.

\bibitem{Rankin2009}
D.W.H. Rankin.
\newblock {\em Handbook of chemistry and physics}.
\newblock Taylor \& Francis, 2009.

\bibitem{Cooke1967}
J.~R. Cooke.
\newblock Some theoretical considerations in stomatal diffusion: A field theory
  approach.
\newblock {\em Acta Biotheor.}, 17(3):95--124, 1967.

\bibitem{Lebedev1965}
N.~N. Lebedev.
\newblock {\em Special functions and their applications}.
\newblock Courier Corporation, 1965.

\bibitem{Leach1993}
A.G. Leach.
\newblock The thermal conductivity of foams. {I}. {M}odels for heat conduction.
\newblock {\em J. Phys. D: Appl. Phys.}, 26(5):733--739, may 1993.

\bibitem{Cantat2013}
I.~Cantat, S.~Cohen-Addad, F.~Elias, F.~Graner, R.~H{\"o}hler, O.~Pitois,
  F.~Rouyer, A.~Saint-Jalmes, R.~Flatman, and S.~Cox.
\newblock {\em Foams: Structure and Dynamics}.
\newblock OUP Oxford, 2013.

\bibitem{Maestro2013}
A.~Maestro, W.~Drenckhan, E.~Rio, and R.~Hohler.
\newblock Liquid dispersions under gravity: volume fraction profile and osmotic
  pressure.
\newblock {\em Soft Matter}, 9:2531--2540, 2013.

\bibitem{Hill1916}
L.~E. Hill, O.~W. Griffith, and M.~Flack.
\newblock {V}. {T}he measurement of the rate of heat-loss at body temperature
  by convection, radiation, and evaporation.
\newblock {\em Philosophical Transactions of the Royal Society of London.
  Series B, Containing Papers of a Biological Character},
  207(335-347):183--220, 1916.

\bibitem{Virtanen2020}
P.~Virtanen, R.~Gommers, T.~E. Oliphant, M.~Haberland, T.~Reddy, D.~Cournapeau,
  E.~Burovski, P.~Peterson, W.~Weckesser, J.~Bright, S.~J. {van der Walt},
  M.~Brett, J.~Wilson, K.~J. Millman, N.~Mayorov, A.~R.~J. Nelson, E.~Jones,
  R.~Kern, E.~Larson, C.~J. Carey, I.~Polat, Y.~Feng, E.~W. Moore,
  J.~{VanderPlas}, D.~Laxalde, J.~Perktold, R.~Cimrman, I.~Henriksen, E.~A.
  Quintero, C.~R. Harris, A.~M. Archibald, A.~H. Ribeiro, F.~Pedregosa, P.~{van
  Mulbregt}, and {SciPy 1.0 Contributors}.
\newblock {{SciPy} 1.0: Fundamental Algorithms for Scientific Computing in
  Python}.
\newblock {\em Nature Methods}, 17:261--272, 2020.

\bibitem{Carrier2003}
V.~Carrier and A.~Colin.
\newblock Coalescence in draining foams.
\newblock {\em Langmuir}, 19(11):4535--4538, 2003.

\bibitem{Roux2022}
A.~Roux, A.~Duchesne, and M.~Baudoin.
\newblock Everlasting bubbles and liquid films resisting drainage, evaporation,
  and nuclei-induced bursting.
\newblock {\em Phys. Rev. Fluids}, 7:L011601, Jan 2022.

\bibitem{Nath2022}
S.~Nath, G.~Ricard, P.~Jin, A.~Bouillant, and D.~Quéré.
\newblock Thermal marangoni bubbles.
\newblock {\em Soft Matter}, 18:7422--7426, 2022.

\end{thebibliography}

\bibliographystyle{unsrt}

\newpage\clearpage


\section*{Supplementary material}

# Supplementary information: Evaporation-induced temperature gradient in a foam column

François Boulogne[1], Emmanuelle Rio[1], and Frédéric Restagno[1]

[1]Université Paris-Saclay, CNRS, Laboratoire de Physique des Solides, 91405, Orsay, France.

## Liquid fraction profile in a stationary regime

Let us consider a monodisperse foam in contact with the soap solution. The bubble radius is denoted $R$ and the foam is made with a liquid of density $\rho$, viscosity $\eta$, and surface tension $\gamma$. We denote the capillary length $\ell_c = \sqrt{\gamma/\rho g}$.

For the sake of simplicity, we choose in this calculation, the origin of the $z$ coordinate at the liquid-foam interface, $z$ being positive in the foam. The liquid fraction $\varphi(z,t)$ has a value $\varphi(0,t) = \varphi_c$ at this liquid-foam interface.

The continuity equation of the liquid phase reads

$$\frac{\partial \varphi}{\partial t} + \frac{\partial u\varphi}{\partial z} = 0, \tag{S1}$$

where $u(z,t)$ is the local velocity of the liquid in the foam along the vertical direction.

**Darcy's law**  The liquid flux, defined as $u_m = u\,\varphi$, can be expressed by Darcy's law:

$$u_m = \frac{\alpha(\varphi)}{\eta}\left(-\rho g + \frac{\gamma}{\delta R}\frac{\partial \varphi^{-1/2}}{\partial z}\right), \tag{S2}$$

where the first term on the rhs accounts for the gravity and the second term is the gradient of capillary pressure. The function $\alpha(\varphi)$ is the foam permeability and $\delta = 1.73$ is a geometric constant (See [1], p. 124).

Assuming that the liquid flow occurs mainly in the Plateau-borders, the permeability can be written as (See [1], p. 124)

$$\alpha(\varphi) = \kappa K_p R^2 \varphi^2, \tag{S3}$$

with $\kappa \approx 0.16$ a geometrical coefficient and $K_p$ the permeability coefficient. For large Boussinesq numbers, *i.e.* immobile interfaces, $K_p = 0.02$ [1].

**Dimensionless differential equation and boundary condition**  We define the dimensionless quantities $\tilde{t}$, $\tilde{\varphi}$ and $\tilde{z}$ as

$$\tilde{t} = \frac{t}{\hat{t}}, \qquad \hat{t} = \frac{\eta \delta^2}{\sqrt{\gamma \rho g}\kappa K_p}; \tag{S4}$$

$$\tilde{\varphi} = \frac{\varphi}{\hat{\varphi}}, \qquad \hat{\varphi} = \frac{\ell_c^2}{R^2 \delta^2}; \tag{S5}$$

$$\tilde{z} = \frac{z}{\hat{z}}, \qquad \hat{z} = \ell_c. \tag{S6}$$

Combining equations S1 and S2, we obtain (See Eq. (3.89) in [1])

$$\frac{\partial \tilde{\varphi}}{\partial \tilde{t}} = \frac{\partial}{\partial \tilde{z}}\left(\tilde{\varphi}^2 + \frac{1}{2}\tilde{\varphi}^{1/2}\frac{\partial \tilde{\varphi}}{\partial \tilde{z}}\right). \tag{S7}$$

At $\tilde{z} = 0$, we have $\tilde{\varphi}(0,t) = \tilde{\varphi}_c = \varphi_c \ell_c^2 / R^2 \delta^2$.

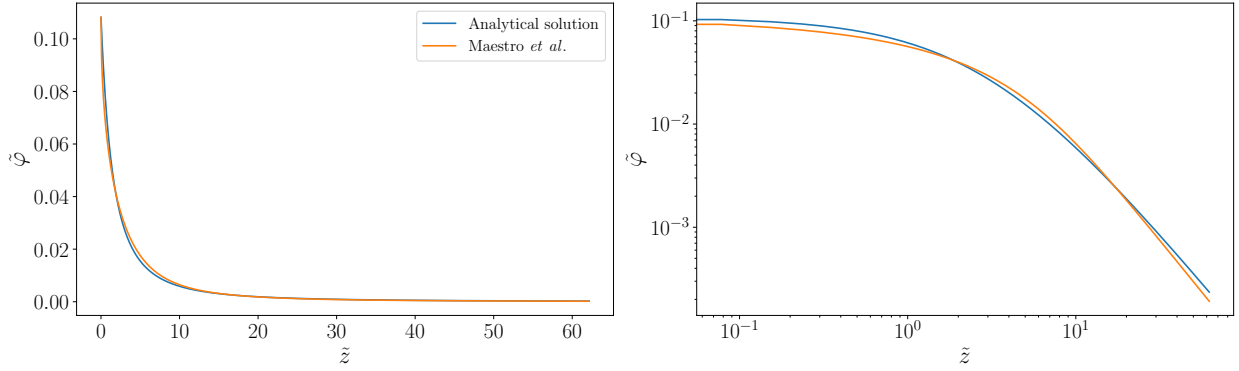

Figure S1: Comparison of the solution proposed by Maestro *et al.* [2] and equation S9.

**Stationary solution** In a stationary regime, no net liquid flux occurs in the foam, *i.e.* $u_m = 0$. This can be justified by considering eq. S1 for $\partial \varphi / \partial t = 0$. Then, it comes that $u \partial \varphi / \partial z + \varphi \partial u / \partial z = 0$. The liquid flow being incompressible, $\partial u / \partial z = 0$. For non-zero liquid fraction gradients, we obtain the condition $u_m = 0$.

In a dimensionless form, this condition can be rewritten

$$\frac{d\tilde{\varphi}}{d\tilde{z}} = -2\tilde{\varphi}^{3/2}, \tag{S8}$$

which can be integrated with the condition $\tilde{\varphi}(\tilde{z} = 0) = \tilde{\varphi}_c$, as

$$\tilde{\varphi} = \left(\tilde{z} + \tilde{\varphi}_c^{-1/2}\right)^{-2}. \tag{S9}$$

**Comparison with the literature** Maestro *et al.* proposed a prediction of the liquid fraction of a foam column, which has been successfully validated experimentally [2]. This prediction, presented in an inverted form $\tilde{z}(\varphi)$ is

$$\tilde{z}(\varphi) = k\left(\sqrt{\varphi_c} - \sqrt{\varphi}\right)\left(3 + \sqrt{\frac{\varphi_c}{\varphi}}\right) + \frac{k}{2}\left(3 - 2\varphi_c - \varphi_c^2\right) \ln\left(\frac{(\sqrt{\varphi} + 1)(\sqrt{\varphi_c} - 1)}{(\sqrt{\varphi} - 1)(\sqrt{\varphi_c} + 1)}\right) \tag{S10}$$

with a numerical prefactor $k = 7.3$.

In figure S1, we show that equation S9 compares well with equation S10. In our work, we used equation S9 for its simplicity.

# References

[1] Cantat I, Cohen-Addad S, Elias F, Graner F, Höhler R, Pitois O, et al. Foams: Structure and Dynamics. OUP Oxford; 2013.

[2] Maestro A, Drenckhan W, Rio E, Hohler R. Liquid dispersions under gravity: volume fraction profile and osmotic pressure. Soft Matter. 2013;9:2531-40.



\end{document}